\documentclass[utf8]{FrontiersinHarvard} 

\usepackage{url,hyperref,lineno,microtype,subcaption}
\usepackage[onehalfspacing]{setspace}

\def\keyFont{\fontsize{8}{11}\helveticabold }
\def\firstAuthorLast{Dobbs {et~al.}} 
\def\Authors{Clare Dobbs\,$^{1,*}$}
\def\Address{$^{1}$School of Physics and Astronomy, University of Exeter, Stocker Road, Exeter, EX4 4QL, UK}
\def\corrAuthor{Corresponding Author}

\def\corrEmail{c.l.dobbs@exeter.ac.uk}

\begin{document}
\onecolumn
\firstpage{1}

\title {2a Results: Galaxy to cloud scales} 

\author[\firstAuthorLast ]{\Authors} 
\address{} 
\correspondance{} 

\extraAuth{}

\maketitle

\begin{abstract}
Simulations from the scales of isolated galaxies to clouds have been instrumental in informing us about molecular cloud formation and evolution. Simulations are able to investigate the roles  of gravity, feedback, turbulence, heating and cooling, and magnetic fields on the physics of the interstellar medium, and star formation.
Compared to simulations of individual clouds, galactic and sub-galactic scale simulations can include larger galactic scale processes such as spiral arms, bars, and larger supernovae bubbles, which may influence star formation. Simulations show cloud properties and lifetimes in broad agreement with observations. Gravity and spiral arms are required to produce more massive GMCs, whilst stellar feedback, likely photoionisation, leads to relatively short cloud lifetimes. On larger scales, supernovae may be more dominant in driving the structure and dynamics, but photoionisation may still have a role. In terms of the dynamics, feedback is probably the main driver of velocity dispersions, but large scale processes such as gravity and spiral arms may also be significant. Magnetic fields are generally found to decrease star formation on galaxy or cloud scales, and simulations are ongoing to study whether clouds are sub or supercritical on different scales in galaxy scale simulations. Simulations on subgalactic scales, or zoom in simulations, allow better resolution of feedback processes, filamentary structure within clouds, and the study of stellar clusters.
\section{}

\tiny
 \keyFont{ \section{Keywords:} keyword, keyword, keyword, keyword, keyword, keyword, keyword, keyword} 
\end{abstract}

\section{Introduction}
Simulations on the scales of isolated galaxies and on kiloparsec scales have allowed us to resolve the multiphase interstellar medium (ISM), molecular clouds, and larger scale feedback processes such as supernovae whilst also including larger scale galactic structure such as spiral arms, bars or galaxy centres.  Here we present a review of the simulations in this field, what they have told us about molecular cloud formation and evolution, star formation, and the role of galactic structure. We start in the introduction with a more historical review of simulations on these scales up to around 2010. The following sections have a stronger focus on work from the last 15 years. Apart from the oldest (i.e. pre 1990) work mentioned, almost all the simulation work cited uses grid or Adaptive Mesh Refinement (AMR), Smoothed Particle Hydrodynamics (SPH), or moving mesh codes. The final section presents conclusions and suggestions for future progress. Moving forward, current progress may involve studying stellar clusters on these scales, and simulating the process of cluster dissolution from its natal gas, and perhaps on larger scales modelling galaxies in a more extragalactic context.

The earliest simulations on galaxy scales focused on understanding the development of galactic structure, and the role of cloud-cloud collisions. For example test particle, and N-body calculations showed the development of spiral or barred structure in galactic discs (e.g. \citealt{Miller1970,Hohl1971,Toomre1972,Sellwood1980,Combes1981,Sellwood1984}). Simulations of molecular or interstellar clouds adopted collisional models with assumptions about cloud growth and destruction to model the resultant mass spectra of clouds, and the influence of a spiral potential on the resultant cloud population \citep{Tomisaka1984,Roberts1984,Combes1985,Roberts1987}. 

Probably the first hydrodynamical simulations of isolated galaxies resembling modern simulations were by \citet{Wada1999} and \citet{Wada2000} who, although they only modelled 2D rather than 3D discs, included a degree of physics surprisingly commensurate with modern day simulations. This included cooling and heating, molecular chemistry, stellar feedback in terms of both winds and supernovae, and test particles sampling the IMF such that insert winds and supernovae are added according to stellar masses and lifetimes. These simulations, and others \citep{Dobbs2008} which included the cold phase of the ISM, allowed smaller clouds to form, and showed the development of larger giant molecular clouds (GMCs) through a combination of gravitational instabilities and cloud-cloud collisions.  Simulations adopting an isothermal warm medium allowed easier predictions with theory. \citet{Elmegreen1983} and others predicted that GMCs form via the development of gravitational instabilities occurring at the most unstable wavelength (fastest growing mode). For a warm isothermal medium this equates to GMCs along spiral arms with characteristic masses of around $10^6-10^7$ M$_{\odot}$ and separations of around 1 kpc, which are reproduced in simulations \citep{Kim2006,Dobbs2008}. 

Some of the earlier hydrodynamical simulations of galaxies also focused on the development of spurs along spiral arms \citep{Kim2002,Chakrabarti2003,Wada2004,Dobbs2006,Kim2006,Shetty2007}. These have been associated with instabilities which include Kelvin Helmholz \citep{Wada2004,Renaud2013}, feathering \citep{Chakrabarti2003,Lee2014}, wiggle \citep{Kim2014}, MRI \citep{Kim2002} and gravitational \citep{Kim2006}, and cloud collisions due to orbit crowding \citep{Dobbs2006}.  In most of these simulations however, the formation of the spurs is independent of how structure is produced. For example if GMCs form, the development of spurs occurs the same way, through the velocity field of the interarm region causing the GMCs to be sheared into large scale filamentary features. The exception is when the instability producing the spurs are associated with resonances in the disc and thereby occur at certain radii in the galaxy \citep{Chakrabarti2003}.

Other simulations focused on the driving of structure in the interstellar medium due to supernova feedback \citep{Rosen1995,Korpi1999,Gazol1999,Wada2000,Wada2001,Slyz2005,Avillez2005,Joung2006}. These simulations, a precursor of the more recent Simulating the lifecycle of molecular Clouds (SILCC) simulations \citep{Walch2015}, typically model a section of galaxy disc and include cooling and heating of the ISM via a cooling curve, stellar feedback and a vertical disc gravitational potential. 

The ISM in these simulations with supernovae feedback is characterised by cold filamentary structures embedded in a warmer medium, and superbubbles generated by supernovae. Cold clouds are able to form even in the absence of self gravity, where supernovae shells collide, though self gravity significantly effects the density PDF \citep{Joung2006}. In the absence of any spiral structure the gas does not form larger clouds, and the clouds tend to be very filamentary. Even when self gravity is included, the ISM in these simulations shows only small scale structure \citep{Wada2001,Jeffreson2021} (for Milky Way or LMC type galaxies, for gas rich low rotation systems which do form massive clumps see e.g. \citealt{Escala2008}).

As simulations started to include most of the key physical processes relevant to molecular cloud formation and evolution, namely gravitational instabilities, thermodynamics of the ISM and stellar feedback, it started to be possible to investigate molecular cloud properties in simulations. 
Including such physics produces a realistic population of clouds whose properties and lifetimes can be evaluated, albeit these may change depending on the details of the feedback, and the inclusion of other processes the most important of which is likely magnetic fields.
 
The past decade or so has seen a greater concentration on the modelling of feedback processes. This is due to a number of factors, including the realisation that stellar feedback is required to create realistic disc galaxies, with the focus on feedback in larger scale simulations leading to interest of the effects of feedback more generally on galactic and ISM evolution. In particular, feedback has been thought to be necessary to produce a low star formation efficiency, and is also required to produce a realistic three phase medium.
Including feedback is complex since there are many different processes to consider, so identifying the role of each has required a large investment of time to incorporate processes into numerical codes, and a large number of simulations, thus feedback has probably been the main focus of simulations in the last decade or so.  

Separating galaxy scale simulations, subgalactic scale simulations and those of individual molecular clouds or colliding flows is difficult, since many processes are investigated across simulations covering multiple scales. Instead references are still included to smaller (molecular cloud) scale simulations and their relation to larger scales considered.

\section{Molecular cloud formation}
The processes by which molecular clouds are thought to form are described in detail in \citet{Dobbs2014}. Here there is a stronger focus on the numerical simulations which demonstrate cloud formation by different mechanisms, and the different conditions under which these mechanisms will dominate cloud formation.

Any instability in the gas, perturbations or development of structure in the gas, will promote thermal instabilities, and lead to denser structures in the cold stable HI phase of the ISM. In the absence of gravity or magnetic fields, such instabilities are the main driver of structure.  
Converging or colliding flows, which could nonetheless result from spiral arms or supernova explosions into surrounding media, allow the accumulation of more significant amounts of thermally unstable gas. Simulations have modelled the thermal instability in converging flows, by modelling smooth colliding flows seeded with a small perturbation, for example a sinusoidal density perturbation \citep{Hennebelle1999,Koyama2002,Heitsch2006}. These simulations show that the thermal instability is able to drive the formation of small clouds.  \citet{Inoue2009} suggest that repeated compression of gas leads to the build up of larger clouds. Alternatively turbulence readily drives the formation of cold structures in the gas \citep{Audit2005}.
With the inclusion of gravity, these clouds collapse to form stars \citep{Vaz2000}. In the presence of a sufficiently strong magnetic field, magnetic pressure can suppress structure formation via thermal instability perpendicular to the field (see review by \citealt{Hennebelle2019} and references therein). Perturbations are expected to be preferentially along the field lines, although with a non uniform field there will be some component of pressure parallel to the field, and Alfv\'{e}n waves can also contribute a  pressure force in this direction \citep{Martin1997,Falceta2003,Pinto2012}. 
Simulations on a galaxy scale show that gas can readily cool when converging flows occur at spiral arms \citep{Dobbs2008}. 

A magnetic medium which is otherwise non-self gravitating, and isothermal, but subject to a vertical gravitational potential, may be subject to Parker instabilities whereby denser gas accumulates in the troughs in the magnetic field. Simulations by \citet{Kim1998} show that the density enhancement due to Parker instabilities alone is only a factor of a few, so not viable to form clouds. However \citet{Kim2000} suggest that Parker instabilities could seed gravitational instabilities, and likewise \citet{Kosinski2006} show that Parker instabilities can induce thermal instabilities and thus form molecular clouds. Parker instabilities combined with cooling lead to the formation of cold dense HI clouds in galaxy scale simulations by \citet{Kortgen2019}.

Once structure occurs in the gas, through cold HI clouds, or molecular clouds, it is possible for cloud-cloud collisions to occur. These are much more likely to occur in spiral arms, where a greater amount of dense gas is present, and where the orbits of the clouds converge \citep{Dobbs2006}. Cloud-cloud collisions are evident in simulations of whole galaxies with cold gas \citep{Dobbs2006,Tasker2008,Skarbinski2023}. In galaxies like the Milky Way, cloud-cloud collisions may typically be fairly gentle events with low Mach numbers (e.g. \citealt{Skarbinski2023}), rather than violent `collisions' and may not generally have much effect on the star formation rate, unless the velocities involved are quite high. However they may at least in some cases gather large volumes of cold gas together into more massive clouds.

With the inclusion of self gravity of the gas, the gas is subject to gravitational instabilities. In simulations with warm gas, Jeans instabilities in the gas are clearly evident, since the gas is smooth and cloud-cloud collisions are minimal, cooling is not present, and as mentioned in the previous section the spacing of the structures in the gas can be equated to the Jeans length \citep{Kim2006,Dobbs2008}. With cooling, the sound speed of the gas is variable in the ISM, allowing gravity to induce structure over the range of wavelengths which are unstable \citep{Elmegreen1989}. 

Finally molecular cloud formation can be induced by colliding flows from supernovae or winds. As mentioned above, supernovae are often assumed to lead to the converging flow set-ups considered in those types of simulations, and are presumed to drive the formation of structure in vertical box models of the ISM. This 
scenario has been modelled most explicitly by \citet{Dawson2015} who performed simulations of two supernovae. The gas is compressed where the supernovae winds collide, becoming thermally unstable similar to the previous models. The morphology of the region is highly structured, ascribed to Rayleigh Taylor instabilities.

Once all the above physics is included in large scale simulations, it is difficult to cleanly distinguish which processes are dominating. It is likely that all the above are producing at least some effect on the gas distribution and the resultant properties of GMCs. However Parker instabilities seem to play a smaller role, or only operate in conjunction with other processes, whilst gravitational instabilities, and cloud-cloud collisions (particularly in spiral arms) seem to be required to produce more massive clouds.

\subsection{ISM properties}
The development and evolution of GMCs in a larger galactic context is interrelated with the wider scale properties of the ISM. The main properties of the ISM that we can check are the amount of gas in different phases, scale heights of phases, and the structure of the components of the ISM. Observationally there are large uncertainties, and significant variations on the amount of gas in different phases, even just considering spiral galaxies. For the Milky Way, observations suggest that molecular gas constitutes around one third of the mass of the ISM \citep{Kennicutt2012}. For the atomic gas, roughly one third lies in each of the cold, unstable and warm phases \citep{McClure2023}. Thus overall around half the gas may lie in the cold molecular or cold atomic phases and the other half in the unstable or warm regime (the mass of the hot ionised component is negligible). Some galaxies with higher surface densities, e.g. M51 will be mostly molecular, whilst there will also be radial variation within galaxies, for example the inner spiral arms or molecular ring in the Milky Way has a higher molecular fraction \citep{Sanders1984,Bronfman1988}.

Most simulations of isolated Milky Way like galaxies have shown fractions of gas in these phases in agreement with the observations \citep{Harfst2006,Dobbs2008,Dobbs2011,Hill2012}. Modelling of dwarf galaxies shows that the cold gas is dominated by HI rather than H$_2$ \citep{Hu2016}.  The amount of gas in different phases will be set mainly by the dust content of the simulated ISM (see Section~4), the chemistry scheme (or explicit cooling curve) used in the code, as well as the resolution (see \citealt{Kim2017}). At solar metallicity, the chemistry, heating and cooling scheme will produce a $10^4$ K component  (warm neutral medium) which starts to transition to the cold phase of the ISM around 0.1 cm$^{-3}$. Once gas reaches 100 cm$^{-3}$ it should be fully in the cold phase unless directly impacted by feedback. This should then lead to realistic gas fractions in the given components, with supernovae feedback producing hotter gas, though if feedback is non-existent (in the presence of self gravity), or too strong then the cold components will be too large or too small respectively. Cosmic rays will also heat the ISM and effect the transition between warm and cold gas \citep{Huang2022,Kim2023}, though \citet{Rathjen2021} find that they only have a modest influence on gas phases and star formation in their simulations of the ISM. Simulations indicate that at large distances above the midplane cosmic rays dominate over the thermal pressure \citep{Wiener2013,Salem2014,Girichidis2018} since supernovae are absent at these scale heights \citep{Girichidis2018}.

Scale heights of the gas are typically larger than would occur solely from the thermal sound speed, rather they reflect the velocity dispersion associated with turbulence in the ISM. So for example the scale height of the cold ISM is around 100 pc, whilst the warm neutral medium the scale height is around 400 pc, the warm or hot ionised medium several 100's parsecs. Simulations with feedback can reproduce these scale heights \citep{Rosen1995,Dobbs2011,Hill2012,Kim2017,Benincasa2020b}. MRI can also increase the scale height of the gas. In simulations by \citet{Piontek2007} the scale height of the cold gas is around 20 or 40 pc, and the unstable medium around 150 pc, so this mechanism alone is probably not sufficiently to account for the observed values. \citet{Hill2012} find that the addition of the magnetic field does not have a significant impact on the scale height of the gas in their simulation, though it will have a moderate effect on the proportion of gas in different phases due to the magnetic pressure leading to more lower density gas.

\subsection{Cloud properties and lifetimes}
\begin{figure}[h!]
\begin{center}
\includegraphics[width=14cm]{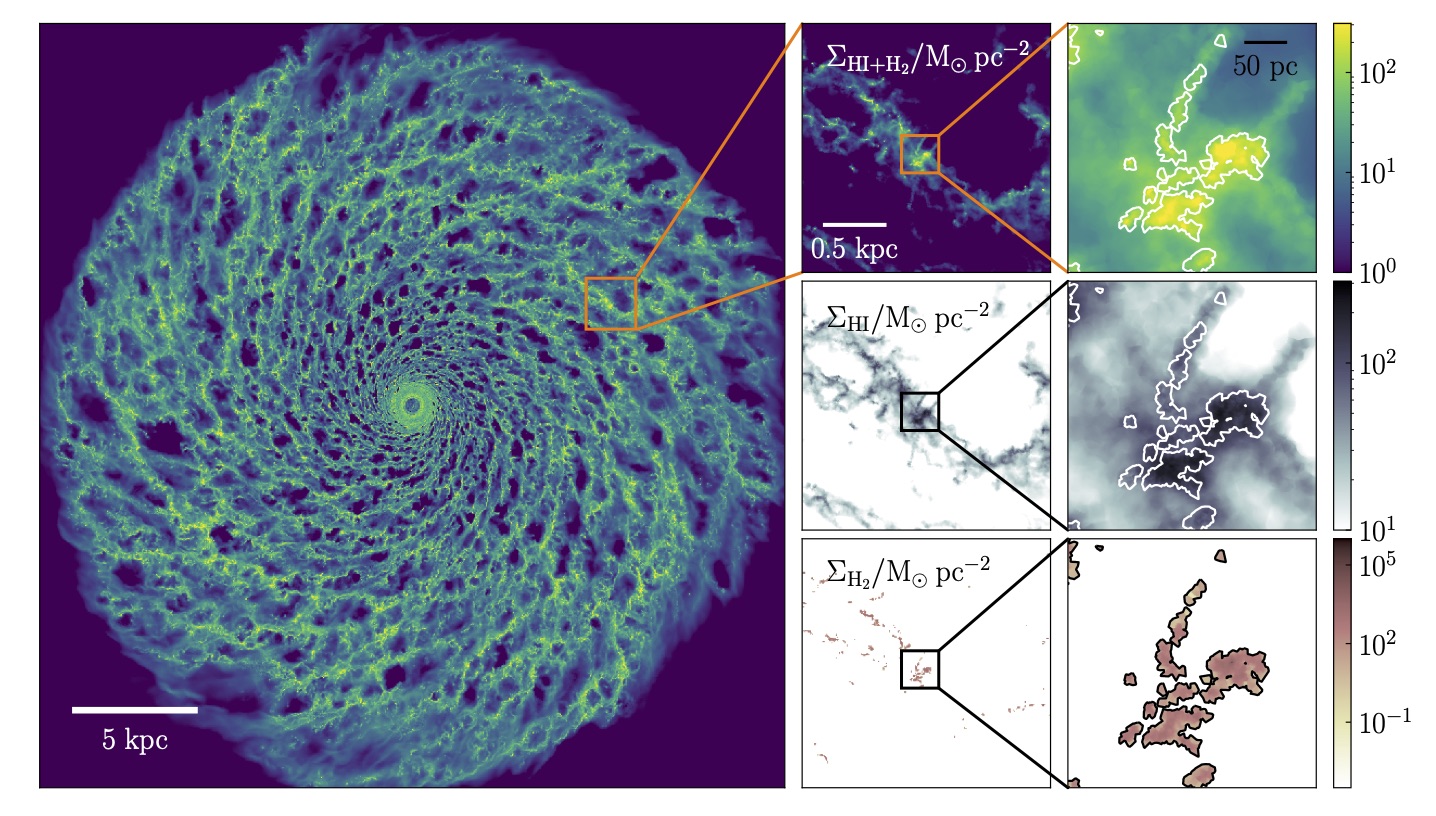}
\end{center}
\caption{Molecular cloud identification in a galaxy simulation by \citet{Jeffreson2020} at a time of 600 Myr. The left hand panel shows the total gas column density, and the right hand panels a 2kpc region seen in total, molecular and atomic column densities, with an example GMC extracted.}\label{fig:1}
\end{figure}

Simulations of isolated galaxies can typically resolve clouds of $10^4-10^7$ M$_{\odot}$, and thus it is possible to start to look at cloud mass spectra, Larson's scaling relations and cloud rotations. Clouds can be selected from simulations using a surface or volume density criterion \citep{Dobbs2006,Tasker2009,Jeffreson2020}, friends of friends algorithm \citep{Dobbs2013,Benincasa2013}, or observationally orientated algorithms such as \textsc{clumpfind} \citep{Grisdale2018} or \textsc{cprops} \citep{Dobbs2019}. Figure~\ref{fig:1} shows a galaxy simulation by \citet{Jeffreson2020}, where the gas is postprocessed to calculate molecular gas fractions and clouds identified at a given surface density contour. 

Simulations obtain cloud mass spectra $dN/dM\propto M^{\alpha}$ where $\alpha$ is typically in the range $-2.5$ to $-1.5$ \citep{Wada2000,Harfst2006,Tasker2009,Khoperskov2013,Jeffreson2020}. Observations find similar results \citep{Fukui2001,Rosolowsky2003,Roman-Duval2010,Rice2016}. \citet{Wada2000} show that a steeper mass spectrum is obtained when including stellar feedback compared to without, and the mass spectrum truncates at around $10^6$ M$_{\odot}$ compared to $10^7$ M$_{\odot}$, indicating that feedback is preventing the formation of more massive clouds. Some results show that without feedback, the mass spectrum is incompatible with observations with too many massive GMCs \citep{Dobbs2011,Grisdale2018}. This is in contradiction to \citet{Tasker2009} and \citet{Wada2000}, where clouds don't seem to require feedback to disperse (rather this is attributed to velocities induced by cloud collisions), presumably due to differences between the numerical codes used. The mass spectrum may vary slightly according to environment, e.g. spiral arms \citep{Colombo2014} or the Outer Galaxy \citep{Padoan2016}, whilst  the slope may depend on the algorithm to find the clouds \citep{Dobbs2019}, and the viewing perspective of the clouds \citep{Khoperskov2016}. Minor differences may occur according to the nature of the spiral arms \citep{Pettitt2020}.

Global disc or spiral arm scale simulations are able to reproduce the Larson's relations \citep{Dobbs2011,Benincasa2013,Falceta2015,Khoperskov2016}, but show quite a lot of scatter compared to at least early Milky Way studies (e.g. the original \citealt{Larson1981} results). They tend to show better correspondence with more recent observational extragalactic studies \citep{Colombo2014,Sun2018}. This could indicate that more recent results are less hampered by interdependencies between different observed cloud properties, the size of the aperture or whether clouds / apertures are used \citep{Spilker2022}, or that there is a greater variation with environment when considering whole galaxies \citep{Rice2016}. The latter is also supported by \citet{Padoan2016} who only consider the Outer Galaxy environment and find a similar degree of scatter in both observations and simulations. 

Molecular clouds are considered bound if their virial prameter
\begin{equation}
\alpha=\frac{5  \sigma^2 R}{GM},
\end{equation}
(e.g. \citealt{Bertoldi1992}) is $<1$, and unbound if $\alpha>2$, with intermediate values being possibly bound or unbound.
Most simulations have shown that molecular cloud population contains both bound and unbound molecular clouds \citep{Tasker2009,Dobbs2011b,Khoperskov2016,Grisdale2018,Lu2020}, again in agreement with observations {\citep{Rosolowsky2007,Sun2018,Sun2020,Evans2021,Duarte2021}. \citet{Dobbs2013} found that the most massive clouds tended to be those which are most gravitationally bound. This is not that surprising since these clouds are the least transient, and also tend to undergo more mergers and cloud-cloud collisions which promote their growth. Numerous processes likely contribute to make GMCs unbound, including collisions, streaming motions, shear and stellar feedback. It is difficult to study these processes in isolation, but at least without stellar feedback, the population of GMCs becomes strongly bound and unrealistic compared to observations \citep{Dobbs2011,Grisdale2018}.  Similar to cloud mass spectra, projection effects may influence the resulting relation of cloud velocity mass radius relations \citep{Shetty2010,Pan2016}. 

Cloud rotations are another diagnostic by which to check whether the simulations are producing GMCs which match the properties of observed GMCs \citep{Phillips1999,Imara2011a,Imara2011b,Braine2018}. AMR simulations can have a tendency to produce GMCs with excessive angular momentum, apparent as spiral shaped rotating clouds, but \citet{Seifried2017} show that with careful restraints on refinement, this issue is avoided. Simulations show that GMCs formed via gravitational instability tend to exhibit prograde rotation (i.e. their internal rotation is the same as that of the galaxy), but cloud-cloud collisions can lead to GMCs rotating in the opposite direction exhibiting retrograde rotation \citep{Dobbs2008,Tasker2009,Jeffreson2020,Aouad2020}. The distributions of prograde and retrograde clouds match well the observational results. \citet{Williamson2014} also find that stellar feedback increases the fraction of retrograde clouds, whilst in the Galactic Centre, shear may determine cloud rotation \citep{Kruijssen2019}. 

Simulations indicate that cloud-cloud collisions are expected to occur reasonably frequently in spiral galaxies. For example \citet{Tasker2009} give a time of $1/5$ of an orbital time between collisions, \citet{Dobbs2015} around 8 Myr. Cloud-cloud collisions in a non-interacting galaxy simulation typically occur at velocities of a few km s$^{-1}$ \citep{Dobbs2015,Skarbinski2023} to around 10 km s$^{-1}$ \citep{Fujimoto2014}, with a maximum velocity of around 20 km s$^{-1}$. This is in line with Milky Way observations, whereby the highest velocity collisions are estimated to be around 15-20 km s$^{-1}$ \citep{Furukawa2009,Fukui2015,Schneider2023}.

Cloud lifetimes are discussed extensively in \citet{Chevance2022}, but an overview of the simulation side is presented here as well. Most simulations appear to produce GMCs with fairly short lifetimes, typically around 4 to 10 Myr \citep{Rogers2013,Dobbs2013,Kim2018,Benincasa2020}, with cloud disruption occurring due to feedback. These lifetimes are at the lower end of most observational studies. \citet{Dobbs2013} find the longest lived clouds associated with the most massive, virialised clouds. \citet{Benincasa2020} see no relation of cloud lifetime with mass, but do similarly see longer lifetimes with lower virial parameter clouds. \citet{Jeffreson2021} find longer lived clouds, with lifetimes of 10-40 Myr. They introduce a momentum feedback scheme which effects the degree of clustering of supernovae and can reduce or increase cloud lifetimes. However even without this scheme clouds would be longer lived compared to the other results, and for example the simulations by \citet{Kim2018} are already high enough resolution to presumably resolve clustering of massive stars. Possibly differences in how clouds are identified between models may contribute to different timescales, as well as any differences in feedback prescriptions. 

\section{Feedback}
Earlier isolated galaxy studies tested whether feedback produces a realistic ISM, and GMC properties in agreement with observations. Driven at least partly by observational studies, the focus of more recent work has moved towards the localised relation between GMCs and star formation, for example the lifetimes of GMCs as discussed above, the time for clusters to move from the embedded to exposed phase, and the most important form of feedback driving these timescales.

Of earlier isolated galaxy simulations with feedback, \citet{Wada2000} include test star particles representing massive stars, which inject energy continuously to represent winds, and deposit thermal energy equivalent to $10^{51}$ ergs each time a supernova occurs. \citet{Tasker2011} include photoelectric heating in the form of a radially dependent heating rate around star particles, which similar to \citet{Wada2000}, are inserted when gas is cold and dense. Both of these works simulate disc galaxies, but with highly flocculent spiral structure only arising in the gas. \citet{Dobbs2011} include a simplified feedback prescription, whereby energy is inserted instantaneously following star formation a snowlplough solution. \citet{Hopkins2011} include feedback as an input of momentum in clumpy regions identified as star-forming. Both \citet{Dobbs2011} and \citet{Hopkins2011} obtain realistic star formation rates regulated by feedback (the former using a galaxy with a spiral potential, the latter using a live stellar disc), though an efficiency parameter for the feedback is required.
In the simulations by \citet{Tasker2011}, the heating does not appear to have a large effect on the morphology of the galaxy, the star formation rate, or the properties of the clouds, although it does seem to affect their rotation. The supernovae schemes in \citet{Wada2000} and \citet{Dobbs2011}, in contrast have a strong impact on the disc, and the resulting clouds. As mentioned above \citet{Wada2000} see the maximum mass truncated with their supernovae plus winds scheme, whilst other simulations see a bimodal mass spectrum dominated by high mass clouds \citep{Dobbs2011,Grisdale2018}. \citet{Hopkins2011} also find models with cooling and self gravity, but no stellar feedback problematic, seeing the formation of very massive clouds, runaway collapse to high densities, high star formation rates and unrealistic galaxy morphologies. The differences are probably two fold. First, \citet{Tasker2011} see shorter lived clouds forming even without feedback, kept in virial equilibrium through cloud-cloud collisions. \citet{Wada2000} are also able to run simulations without feedback without seeing a build up of high mass clouds. Secondly the photoionising heating scheme is probably gentler than a large deposit of energy more characteristic of the supernovae schemes. 

In more recent simulations, the emphasis has been to consider the timescale of different feedback processes, including early (e.g. photoionisation, winds) and late (e.g. supernovae) feedback. \citet{Iffrig2015} find that the location of supernovae, whether they are randomly distributed in the ISM when they occur, or distributed in dense gas, makes a large difference on the galactic structure and star formation rate. \citet{Gatto2015} also find that if supernovae occur in dense gas this leads to both too little molecular gas and too much hot gas. Many smaller scale simulations over the past decade have pointed towards ionisation being effective in dispersing clouds before supernovae occur (see Figure~\ref{fig:2}) on timescales of a few or several Myr \citep{Dale2005,Sales2014,Geen2016,Gavagnin2017,Peters2017,Kim2018,Haid2019,Ali2018,Lucas2020,Gonzalez2020,Fukushima2020,Bending2022,Dobbs2022a}. 

\begin{figure}[h!]
\begin{center}
\includegraphics[width=13cm]{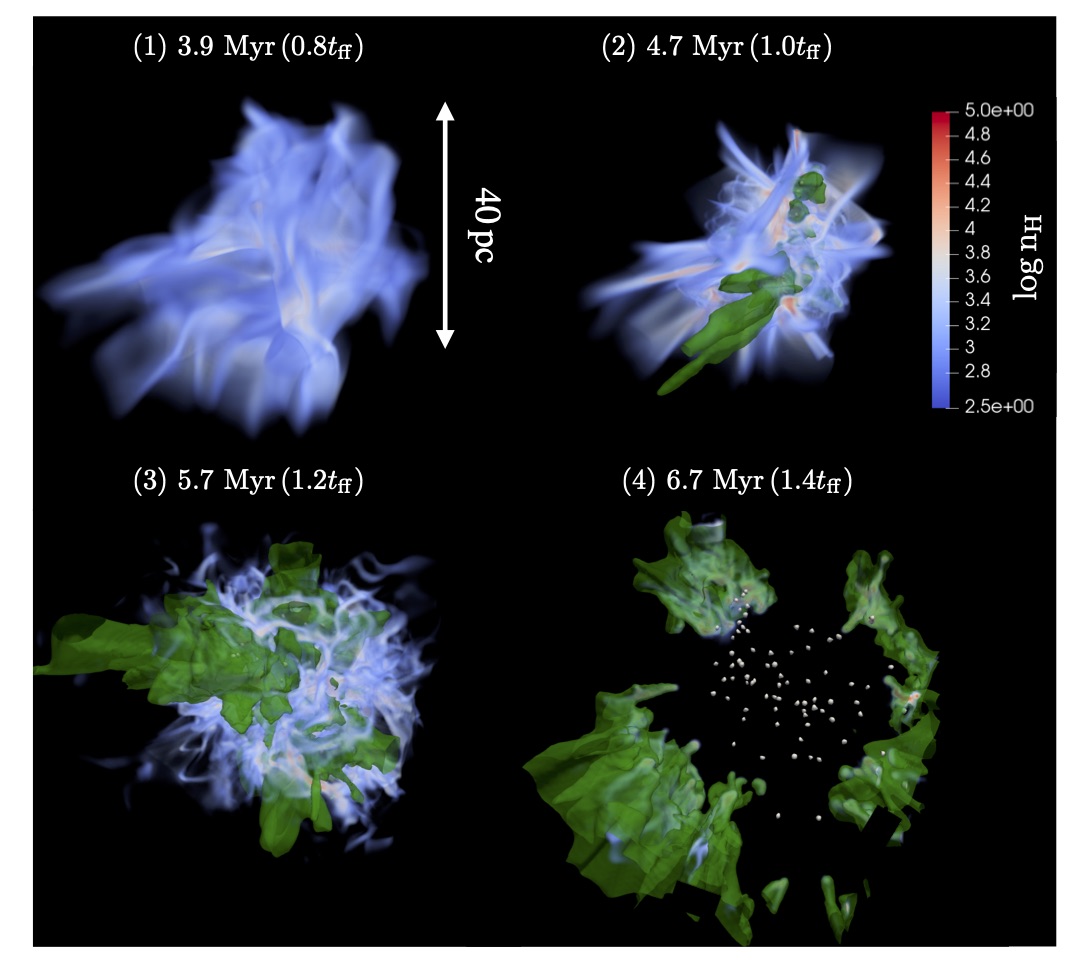}
\end{center}
\caption{A 3D volume rendering of the density field for a $10^5$ M$_{\odot}$, solar metallicity cloud subject to ionising radiation is shown at four different times, taken from \citet{Fukushima2020}. The ionisation fronts are shown in green, and the white dots in the final panel represent the positions of the star cluster particles. The length scale for each panel is 40 pc.}\label{fig:2}
\end{figure}

Ionisation is more difficult to resolve on larger scales, and typically (though see \citealt{Benincasa2020b}) simulations heat gas to $10^4$ K within a Stromgen radius \citep{Marinacci2019,Baba2017,Jeffreson2021}. Ionisation has been included in whole galaxy simulations but there is not a clear consensus yet on the role of ionisation on larger scales. \citet{Rosdahl2015} find that ionisation is gentler and less effective than supernovae whilst \citet{Benincasa2020b} find supernovae are required in addition to ionisation to regulate the star formation rate in their simulations. \citet{Hu2016} find supernovae are dominant in simulations of dwarf galaxies. However \citet{Jeffreson2021} highlight the importance of photoionisation at earlier times (compared to supernovae) to regulate cloud lifetimes, whilst \citet{Stinson2013} find early feedback is important for cosmological simulations. It seems likely that both play a role, e.g. photoionisation initially creates cavities and shapes the gas but this is further accentuated by supernovae \citep{Rogers2013, Grudic2022, Herrington2023}. The current literature seems to find that on larger scales supernovae and ionisation are dominant compared to winds \citep{Marinacci2019,Ali2022} and radiation pressure \citep{Marinacci2019}.  Photoionisation may have a greater role for simulations whereby gravitational collapse to high densities occurs without early feedback. 

The role of feedback is likely to be complex and will vary according to length and time scales, and not necessarily always to lower the star formation rate. On larger galaxy scales, and over 10s or 100s Myr timescales, feedback is required to prevent the accumulation of most of the gas into massive long lived clouds which would turn most of their gas into stars. Instead feedback keeps clouds in a continuous state of flux, their lifetimes relatively short, retains high velocity dispersions in the ISM and returns gas to lower, warmer phases of the interstellar medium, keeping the star formation moderate. Other processes, such as gravity, spiral shocks and cloud-cloud collisions (see next section) may drive turbulence, which similarly contributes to the galaxy wide velocity dispersion but as discussed in Section~2.2, simulations are somewhat divided as to whether these are sufficient to maintain a population of turbulent, short-lived clouds. Cloud-cloud collisions are usually found to produce similar or higher star formation rates compared to clouds in isolation \citep{Wu2017,Dobbs2020,Liow2020,Tanvir2020,Hunter2023}). 
On intermediate scale however, feedback may have more of a triggering effect. This is not always so evident in numerical simulations, since if they typically only model an isolated cloud, the feedback simply escapes into a vacuum. Observers and theorists have nevertheless long suggested that feedback triggers star formation (e.g. \citealt{Elmegreen1977,McCray1987,Tenorio1987,Elmegreen2002}). \citet{Herrington2023} find that when including supernovae from a previous generation of stars, the star formation rate is increased compared to not having a preexisting population. This is due to the supernovae (and to a lesser extent ionisation) compressing structure in the gas to higher densities, and thereby forming stars. Comparing this to the picture by \citet{Inutsuka2015}, whereby molecular clouds are envisioned to lie at the edges of expanding shells, it may be that most star formation occurs in this way. In equilibrium, the feedback simply converts a continuous steady amount of the ISM into star forming gas. On smaller, core scales, feedback again appears to reduce star formation (e.g. \citealt{Guszejnov2022}).

The impact of the magnetic field on the larger scale effects of feedback has been investigated by a few authors. Using both theoretical arguments and simulations, \citet{Leao2009} find that the size of supernovae remnants is reduced in the presence of a strong magnetic field, so predict that supernovae may have a limiting contribution to the effectiveness of feedback in shaping the ISM and triggering new star formation. \citet{Iffrig2015} also run simulations of supernovae occurring in a magnetic medium, and find that supernovae combined with a magnetic field are better able to drive the velocity dispersion of the gas. Similar results are seen in simulations of a vertically stratified section of disc \citep{Iffrig2017}, and comparing the resulting density distributions (their Fig.~1), the inclusion of a magnetic field actually appears to produce clearer cavities.

The role of feedback in shaping the structure of the ISM on galaxy scales is apparent in for example the observations of holes in HI \citep{Bagetakos2011} or more recently PHANGS maps of galaxies \citep{Watkins2023a,Watkins2023b}. The holes are widespread and a substantial component by area. Cavities can form simply from the gaps between sheared molecular clouds, however such cavities would be strongly sheared in the interarm regions (see e.g. shells subjects to shear modelled by \citet{Tenorio1987}). Whereas for example in \citet{Watkins2023a}, although highly elliptical bubbles are associated with dynamical creation or older bubbles, many more circular bubbles are also present.  
\citet{Watkins2023b} also show that the sizes and expansions of these features is consistent with having been caused by supernovae from the underlying young stellar population. Observations show the formation of molecular clouds on the edges of superbubble regions  \citep{Tanaka2007,Dawson2011,Dawson2013,Sano2018,Bialy2021}.  More locally, nearby star forming regions appear to be shaped by feedback, and exhibit age spreads consistent with a sequence of star formation occurring, then triggering either new structures or collapse in preexisting dense or molecular clouds \citep{Oey2005,Grossschedl2021,Zucker2022,Miret2022}.

\subsection{Turbulent driving}
Simulations without feedback have suggested that turbulence may be driven by gravitational instabilities \citep{Wada1999,Wada2002,Tasker2009}, MRI \citep{Kim2003} or shear \citep{Fleck1981,Meidt2018}. This might be more relevant in some environments, e.g. dwarf galaxies or low star forming galaxies \citep{Wada2002,Agertz2009}, or the outer parts of galaxies, where feedback is minimal. Thermal instability has also been suggested, which can induce velocities of a few km s$^{-1}$ \citep{Koyama2002}. Spiral arms are also able to induce velocity dispersions in the spiral arms, both from the shock \citep{Kim2006b,Dobbs2007}, and tidal forces acting over larger scales along the spiral arm \citep{Falceta2015}. Most consensus however points towards supernovae driving velocity dispersions of several km s$^{-1}$ up to $100-200$ pc scales \citep{Slyz2005,Dib2006,Joung2006,Dobbs2011,Hill2012,Gent2013,Lu2020,Herrington2023}.  There is still some debate though whether feedback alone drives turbulence, and whether feedback or gravity is the source of the shape of the velocity power law spectrum. Based on modelling sections of the ISM, \citet{Seifried2018} suggest that supernovae may not be frequent enough to maintain velocity dispersions at large distances from clouds. Supernovae, or other feedback from massive stars, may also not be able to explain turbulence deeper within molecular clouds which do not contain massive stars. \citet{Ejdetjarn2022} find in disc galaxy simulations that the velocities reached are the same, due to gravity, regardless of feedback. Most simulations however find a larger velocity dispersion when supernovae are included and that the velocity dispersion increases as a function of how many supernovae occur. \citet{Colman2022} drive turbulence explicitly on given scales as may be typified by feedback processes operating on particular distance scales, but find that feedback alone is insufficient to explain the observed power spectrum particular at the largest scales, whilst \citet{Smith2022} similarly find that turbulence needs to be driven across multiple scales to produce realistic star clusters. \citet{Grisdale2017} suppose that a combination of gravity, shear and feedback drive the turbulent power spectrum. \citet{Bournaud2010} suggest that gravitational instabilities produce the velocity power law (see also \citealt{Fensch2023}) but feedback is essential to transform dense gas to the diffuse phase, and replenish the low density, high velocity dispersion gas covering larger scales. 

On cloud scales, photoionisation has been found to produce a Kolmogorov type spectrum by \citet{Boneberg2015}, but it is not clear how much photoionisation drives turbulence on galaxy scales (e.g. \citealt{vandenbroucke2019}). \citet{Herrington2023} find that supernovae drive larger velocity dispersions compared to photoionisation. \citet{Sartorio2021} also find that if turbulence is already driven by a larger scale process, ionisation from a nearby source may have limited impact on the velocity and density fields.

\section{Chemistry}
In terms of chemistry, isolated galaxy studies have concentrated on including molecular hydrogen and CO formation in their models. Many of these studies have investigated the formation of molecular hydrogen in different environments, and whether such clouds could be observed in CO. But also a second reason for the inclusion of molecular hydrogen formation has been the observation that the Kennicutt Schmidt relation links better to molecular hydrogen than atomic hydrogen or total gas density, so some simulations have used molecular hydrogen density as a threshold for forming stars rather than total density.

The formation of molecular hydrogen and CO are governed by basic equations, 
\begin{equation}
\frac{dn(H_2)}{dt}=R_{gr}(T) n \, n(H)-(\zeta_{cr}+\zeta_{diss}(N(H_2),A_V)n(H_2)
\end{equation}
and
\begin{equation}
\frac{dn(CO)}{dt}=k_0 n(C^+) n \beta(n(H_2),\tau_{UV})-\Gamma_{CO}(\tau_{UV}) n(CO)
\end{equation}
see e.g. \citet{Hollenbach1971,Nelson1997,Bergin2004}. Here, $n$ represent number densities, $N$ is column density, $R_{gr}$ is the formation rate on grains, $\zeta_{cr}$ and $\zeta_{diss}$ are the cosmic ray and photodissociation rates respectively, $k_0$ is a constant, and $\tau$ the optical depth. $\beta$ and $\Gamma$ are functions which additionally depend on the O$_{\rm{I}}$ abundance and the average interstellar radiation field \citep{Habing1968}. In terms of the properties of the gas, the formation of molecular hydrogen is dependent on density squared and temperature, so even though higher temperatures are more conducive to forming H$_2$, the high density of cold gas means that molecules preferentially form in cold dense regions. As illustrated in Figure~14 of \citet{Dobbs2008}, molecular hydrogen formation and destruction versus time exhibits a hysteresis loop. As density increases, H$_2$ fraction increases as molecular hydrogen forms. However once formed, gas can remain molecular at lower densities due to self shielding. Molecular hydrogen formation will also depend on the amount of dust, and the properties of dust grains (see below), which are encapsulated in the $R_{gr}$ parameter.
Simulations on galaxy scales have showed densities, metallicities and temperatures at which H$_2$ forms \citep{Pelupessy2006,Dobbs2006,Dobbs2008,Gnedin2009,Richings2014,Jeffreson2020,Bellomi2020} and that the compression of cold gas in spiral arms or by turbulence is sufficient to form H$_2$. The timescale for gas to move from being atomic to obtaining modest molecular fractions is found to be of order Myr when the gas is compressed by spiral arms of turbulence \citep{Glover2007,Dobbs2008}. 

\citet{Robertson2008} investigated the Kennicutt Schmidt relation and found better agreement with a $n=1.4$ power law when using the molecular gas surface density rather than total gas surface density, which gives a cut off at low densities (see also \citealt{Monaco2012}). Alternatively the H$_2$ surface density can be used when the Kennicutt Schmidt relation is used to input the star formation rate in simulations, in part naturally reducing the star formation efficiency \citep{Lagos2011,Kuhlen2012,Sillero2021,Valentini2023}. By combining with a radiative transfer code, synthetic HI, HISA, CO and also CII maps of galaxies, or sections of galaxies can be produced \citep{Dib2005,Acreman2010,Duarte-Cabral2015,Heiner2015,Seifried2020,Ebagezio2022}.

Simulations with chemistry can also be used to determine some measurements which are observationally challenging, e.g. the fraction of dark molecular gas (see Figure~\ref{fig:3}) , the X factor converting from CO intensities to molecular masses, and variation of the X factor with environment. Simulations indicate that there are significant amounts of dark H$_2$ \citep{Papadopoulos2002,Dobbs2008,Pelupessy2009,Smith2014,Glover2016,Gong2018,Seifried2020} that would not be observed with CO. The X factor has been found to vary with densities, velocity linewidths, star formation rate at high densities, metallicity \citep{Shetty2011,Clark2015} and cosmic ray ionisation rate \citep{Penaloza2018,Bisbas2021}. Seemingly in contrast to \citet{Clark2015}, \citet{Bisbas2021} do not find a variation with FUV, though this could be because they don't consider the same high densities as \citet{Clark2015}. 

\begin{figure}[h!]
\begin{center}
\includegraphics[width=17.5cm]{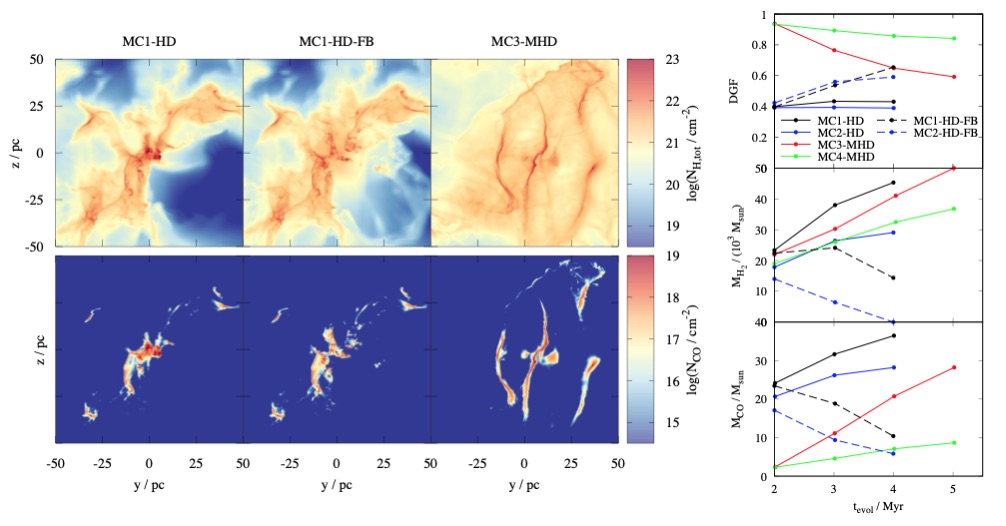}
\end{center}
\caption{The total gas column density (upper left), and CO column density (lower left) are shown from simulations of molecular clouds from \citet{Seifried2020}. The simulations include feedback (centre) and MHD (right). Plots showing the variation of mass of the dark molecular gas fraction, hydrogen and CO are shown on the right.}\label{fig:3}
\end{figure}

Uncertainties in determining molecular densities in simulations include the efficiency of H$_2$ formation on grains, self shielding and the UV field, the amount of feedback and the resolution of the simulations. The efficiency of H$_2$ formation on grains may depend on the grain type, and grain sizes which in turn depend on environment and grain growth, \citep{Smoluchowski1981,Tielens1982,Snow1983,Caselli1998,Biham1998,Biham2001,Cuppen2006,Perets2007}. The efficiency will also depend on how quickly hydrogen atoms combine on the surface of grains, the rate of diffusion. This can be determined through numerical calculations and laboratory experiments  \citep{Duley1996,Ruffle2000,Hincelin2015}. However there is still some uncertainty of what this quantity will be in astrophysical environments. In particular there is found to be significant dependence on the temperature  \citep{Cazaux2004, LeBourlot2012,Iqbal2012,Grieco2023}. 

Self shielding of the molecular hydrogen can be calculated for example using a typical distance to massive stars \citep{Dobbs2006,Khoperskov2013}, or the Jeans length \citep{Glover2015}, but ideally a ray tracing method or tree based method like treecol can be used to determine a better measure of the column density for self shielding \citep{Clark2012,Hartwig2015}. The feedback scheme will also affect the amount of molecular hydrogen. So for example, the amount of H$_2$ will depend both on the level of feedback, and whether feedback quickly disperses a cloud or not \citep{Walch2015}. Finally, although cosmological simulations are keen to include H$_2$ and CO evolution, the results from galaxy and smaller scale simulations, as well as tests by \citet{Krumholz2011b}, suggest that resolving these species accurately requires quite high resolution. \citet{Duarte-Cabral2016} find that subgalaxy simulations with subparsec resolution achieve higher fractions of H$_2$ compared to lower resolution whole galaxy simulations. \citet{Nickerson2019} perform different resolution galaxy simulations and find that  the H$_2$ content is not converged, though their 6 pc resolution simulation produces H$_2$ fractions in agreement with observations. \citet{Gong2018} find that 2 pc resolution is required for convergence of H$_2$, whilst CO is still not converged at 1 pc resolution. \citet{Joshi2019} suggest even higher resolution of 0.2 pc for H$_2$. CO may require a resolution of around 0.1 pc \citep{Joshi2019,Borchert2022}. \citet{Jeffreson2020} suggest the resolution required may also be sensitive to the calculation of self shielding, but still note high resolution is required.

\section{Magnetic fields}
A review of the role of magnetic fields on molecular cloud scales is presented in \citet{Hennebelle2019}, whilst \citet{Pattle2022} review observations and simulations again on molecular cloud scales and smaller.  
\begin{figure}[h!]
\begin{center}
\includegraphics[width=17.5cm]{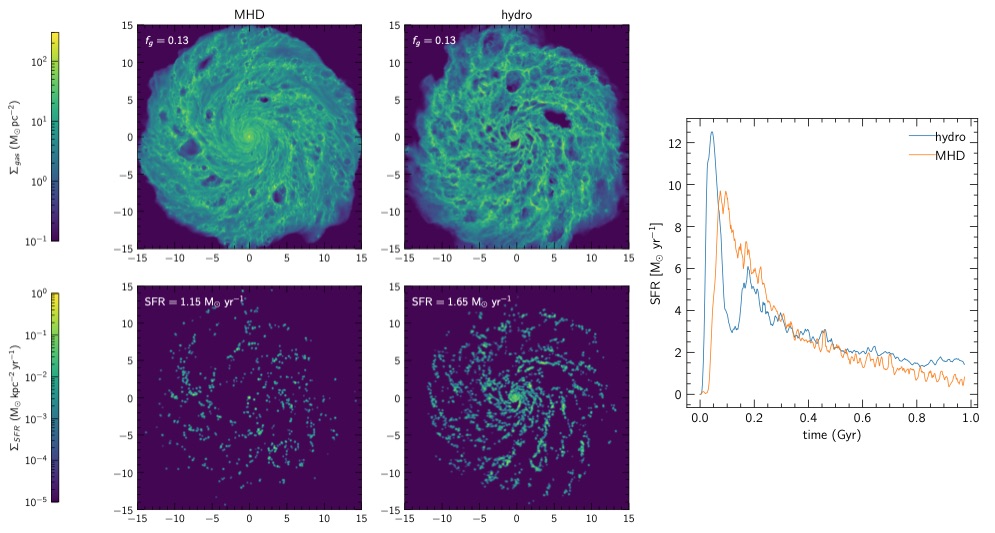}
\end{center}
\caption{Galaxy simulations from \citet{Wibking2023} with and without magnetic fields are shown at a time of 1 Gyr. The upper panels show the density, and the lower panels show the star formation rate per unit area. The right panel shows the star formation rate over time for the two simulations.}\label{fig:4}
\end{figure}

Generally, aside from the possibility of inducing Parker instabilities as discussed above, magnetic fields are thought to reduce and / or slow down star formation in galaxies and molecular clouds, if indeed they have an effect. At the simplest level, magnetic fields exhibit a pressure which on galaxy scales smoothes out structure induced by other processes such as spiral arms and self gravity \citep{Dobbs2008m,Hill2012,Schmidt2013}. On cloud scales, the magnetic field may be so strong that the molecular cloud is subcritical, and the magnetic force is stronger than the gravitational force such that the cloud is stabilised against collapse. The condition for this to occur is that the ratio of the cloud mass to the critical mass M$_{crit}$ is less than 1, ie. 
\begin{equation}
\frac{M}{M_{crit}} \sim 0.17 \frac{\phi_B}{G^{1/2}} \sim 7.6 \times 10^{-21} \frac{N(H_2)}{B} <1,
\end{equation}
where $\phi_B=B \pi R^2$ and the first expression is for a uniform spherical cloud, and the second can be applied to line of sight Zeeman measurements \citep{Crutcher2004,Krumholz2011}. \citet{Crutcher2010} estimate that most clouds are supercritical, and find a change from the magnetic field being constant, to having a $\rho^{2/3}$ dependence at $n\sim 300$ cm$^{-3}$, which they suggest may be where clouds start being gravitationally dominated. Recent simulations by \citet{Auddy2022} of turbulent magnetic ISM are able to reproduce a change in slope, which they find is due to transitioning from sub to supercriticality, and is also dependent on the Alfv\'en Mach number of the turbulence.

Star formation can commence in subcritical clouds via ambipolar diffusion and / or through accretion of mass onto the cloud which could push the mass over the the critical mass.  Simulations of molecular cloud formation have studied whether indeed the clouds formed in the simulations are supercritical, and if so whether they undergo a transition from super to subcritical.  \citet{Kortgen2015} investigate this problem in simulations of driven turbulence and find that above certain field strengths, the clouds are subcritical and it is difficult for star formation to proceed. Likewise \citet{Ostriker1999} find that subcritical clouds do not collapse. However in galaxy scale simulations, accretion onto molecular clouds readily lead them to become supercritical and thereby collapse \citep{Kortgen2018}. Interestingly \citet{Hu2023} suggest that most clouds, even HI clouds, are supercritical, and simply measurements of subcritical clouds are due to observational biases. On galaxy scales, numerous large scale simulations have shown that magnetic fields suppress the star formation rate by a factor of 2 or 3 \citep{Vazquez2011,Kim2021,Wibking2023} (see Figure~\ref{fig:4}). The main exception is recent work by \citet{Whitworth2023}, who model dwarf galaxies and find the magnetic field does not effect the star formation rate.

The effects of magnetic fields may also have a directional dependence. This tends to be the case for relatively strong fields, as for weaker fields the field has less impact on the gas. As noted in \citet{Hennebelle2019}, magnetic fields will effect gas differently parallel or perpendicular to the field. Consequently, a strongly magnetised medium may allow the formation of filaments, but be non-conducive to forming more spherical clouds \citep{Heitsch2009,Inoue2009b,Hennebelle2013,Wareing2016}. Observationally magnetic fields are found to be aligned with density structures in the ISM, but at the highest densities, fields become orientated perpendicular to dense filaments \citep{McClure2006,Alves2008,Heyer2012,Clark2014,Planck2016}. The change in direction has been related to the transition to gravitationally dominated structures \citep{Soler2013,Chen2020,Girichidis2021,Barreto-Mota2021}, though it is worth noting that such a change in orientation has been seen even in simulations without self gravity \citep{Gazol2021}. 
The behaviour of gas at shocks will also differ according to magnetic field orientation. Again for weak fields, the orientation makes little difference. However for strong fields, the magnetic field will be amplified if the field is parallel to the shock. Consequently, if the field is parallel to the shock, when the field is amplified, the gas becomes magnetically supported and star formation is prevented. \citet{Dobbs2021} find that the star formation is delayed rather than prevented, but the morphology of the region is somewhat different. Another consequence is that lower density shocked regions are likely to occur when the field is parallel to the shock, and higher density regions when the field is perpendicular, similar to observations. 

It was previously supposed that magnetic fields might prevent dissipation of turbulence, and hence prolong star formation (e.g. \citealt{Shu1987}). However simulations find that once clouds are supercritical, the timescale for collapse with magnetic fields is not significantly longer than the unmagnetised case \citep{MacLow1998,Ostriker1999}, with the proviso that locally there may be more variation due to directional dependence on the influence of the magnetic field. As such, it may be that magnetic fields keep mass critically supported, rather than slowing down the process of star formation substantially \citep{Vazquez2011}. 

A caveat with both simulations of turbulence and magnetic fields, is that the resolution of the simulations is insufficient to model the full range of Reynolds numbers \citep{Elmegreen2004}, or the range of Prandtl numbers \citep{Nixon2019}, expected for the ISM.   

\section{Environment}
So far we have considered different physical processes that could largely be considered independently of galactic environment. However a galaxy contains multiple environments in which different processes may dominate, and which exhibit different characteristics. Isolated galaxy simulations have the advantage of being able to model different regions simultaneously.

One of the longstanding questions in star formation is whether spiral arms trigger star formation, or simply gather gas, which would anyway be forming stars, together. Spiral arms represent regions of lower shear, and the gas may shock as it traverses the minimum of the spiral potential \citep{Roberts1969}. If the gas is relatively uniform low density, and undergoes a shock at the spiral arms, then it may well reach densities high enough to induce molecule formation and subsequently star formation. 
So for example, in simulations without self gravity, molecular gas is predominantly situated in the spiral arms, but atomic between spiral arms \citep{Dobbs2006,Dobbs2008,Kim2008,Kim2010}, and thus the spiral arms are the dominant mechanism driving molecular, or dense gas formation.
However if the gas already lies in relatively dense clouds, or filaments, then the spiral arms may simply act to gather these clouds and filaments together. 
In later simulations which include self gravity, spiral arms and feedback \citep{Dobbs2011,Pettitt2017,Kim2020,Tress2020}, there is little difference in the star formation rate with and without spiral arms, suggesting that spiral arms do not have a big impact on star formation.  As shown in \citet{Duarte2017} gas entering the spiral arms already includes giant molecular filaments, sheared into their elongated shapes during the inter-arm region passage. The main impact seems to be the formation of a few more massive clouds when spiral arms are included, which though they can have a disproportionate effect on the star formation rate, do not hugely increase star formation \citep{Dobbs2013,Dobbs2017}.

The outer regions of galaxies are typically associated with less star formation, including less massive star formation. \citet{Schaye2004} predicted that the edge of the cold component of galaxies coincides with the range of star formation in the disc. Inspired by GALEX observations of extended massive star formation, beyond the readily observable galaxy discs, \citet{Bush2010} showed in simulations that spiral arms out to large radii could allow star formation to proceed at the outskirts of galaxies. Likewise \citet{Smith2023} find cold gas forming in spiral arms at the edge of galaxies. As mentioned above, other means of driving turbulence may be required at the outer galaxy rather than relying on massive stars, though \citet{Dib2006} suggest supernovae may be sufficient depending on the star formation efficiency. Also, galaxies are not in isolation, and accretion of gas may promote both turbulence and star formation at the edge of galaxy discs. A few simulations have shown that the impact of high velocity clouds can drive turbulence and possibly induce a mini-starburst \citep{Santillan2007,Alig2018}, whilst more generally, the accretion rate onto galaxies may be sufficient to drive turbulence, and allow star formation to continue at a steady rate \citep{Klessen2010,Burkert2017}. 

\begin{figure}[h!]
\begin{center}
\includegraphics[width=12cm]{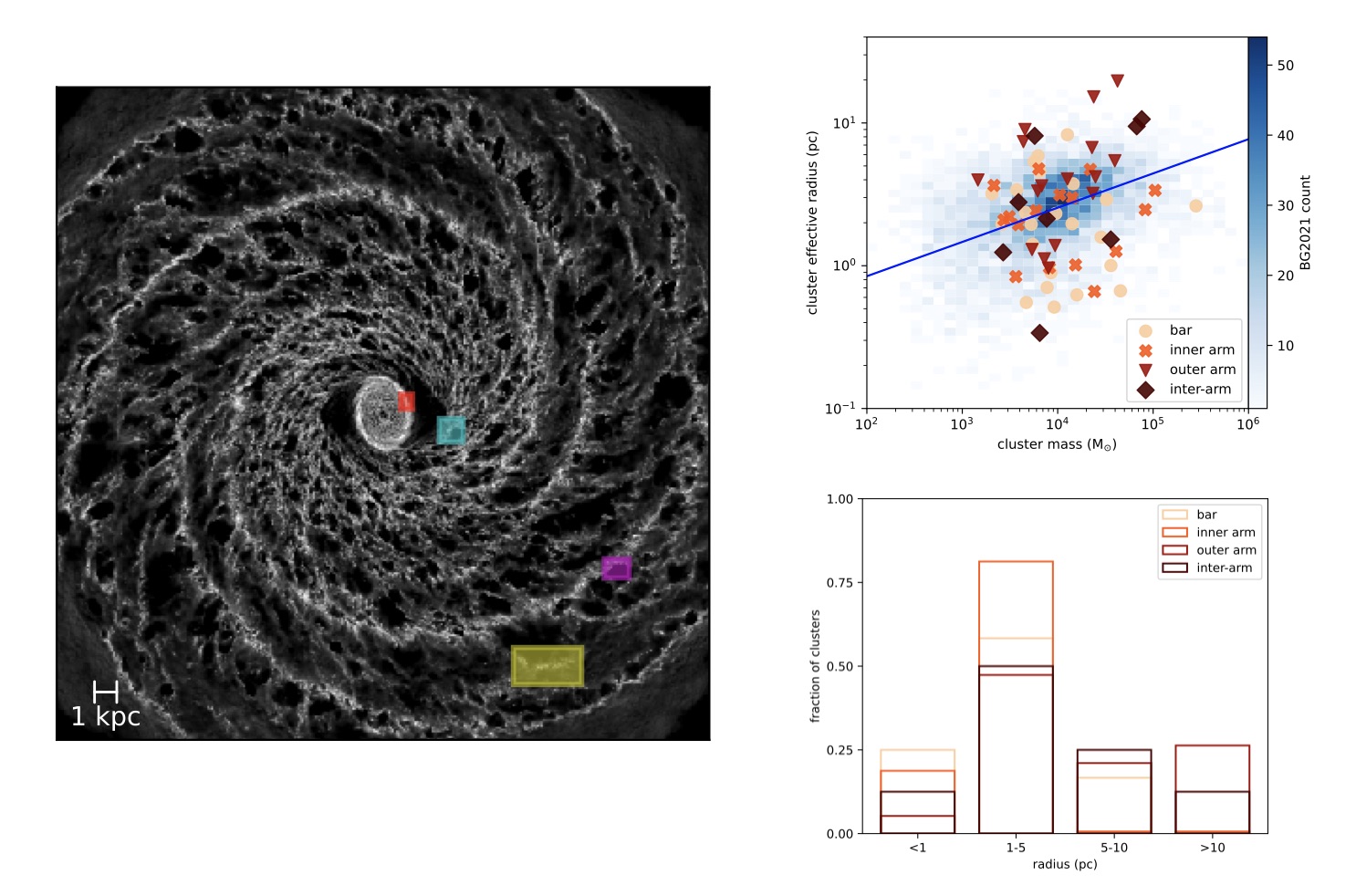}
\end{center}
\caption{Regions from a barred galaxy simulation which are resimulated at higher resolution are shown in the left panel, from \citet{Ali2023}. The right panels show the variation of cluster properties with environment, in particular the outer galaxy and inter-arm reigons produce larger radius clusters, whereas the bar and inner spiral arm produce more compact clusters.}\label{fig:5}
\end{figure}

Similarly to the spiral arms, simulations of the bar region also suggest that cloud mergers are more frequent leading to the formation of more massive clouds \citep{Renaud2013,Fujimoto2014}.  \citet{Renaud2013} also highlight the role of resonances in channeling gas to the ends of the bar to form massive clouds and clusters \citep{Renaud2015b,Ali2023} (Figure~\ref{fig:5}). Most work on galaxy centres has focused on our on Galactic Centre. In simulations of the Galactic Centre, the bar region also channels gas into the galactic centre \citep{Shin2017,Tress2020,Sormani2018,Sormani2019,Moon2021}, fuelling the nuclear ring. Simulations predict that the fuelling of gas leads to collisions with very high velocities, up to $\sim$ 200 km s$^{-1}$ \citep{Sormani2019,Hatchfield2021}. Observationally, molecular clouds at the Galactic Centre show lower star formation rates than their disc equivalents \citep{Longmore2013} so simulations of the Galactic Centre have been used to try and understand why this is. \citet{Bertram2015} perform simulations of molecular clouds subject to conditions at the Galactic Centre, including a strong radiation field representative of the Galactic Centre, but find that star formation still occurs within a gravitational free fall time, and that the amount of star formation is higher than observed. 
However by including the orbit of the clouds at the Galactic Centre, \citet{Kruijssen2019} find that shear and torque are also important conditions needed to better reproduce the observed properties of the clouds, and that the tidal field at the Galactic Centre is required \citep{Dale2019}.

The above simulations of the Galactic Centre aim to model the central part of the Milky Way, but otherwise the above calculations largely model generic Milky Way like spiral or barred spiral galaxies. An alternative approach is to try and model specific galaxies, usually by running many lower resolution, or N-body simulations to determine a good match. This has the advantage that a more robust, and absolute comparison with observational properties can be made since the properties of the simulated and observed galaxy, such as surface densities, rotation curve, spiral pattern roughly match. Quite a few studies have attempted to find a match to the Milky Way by trying to reproduce the CO map of the Galaxy \citep{Wada1994,Englmaier1999,Fux1999,Bissantz2003,Rodriguez2008,Baba2009,Khoperskov2013,Pettitt2014,Pettitt2015,Li2016,Li2022}. The CO map has the advantage that it traces fairly narrow features compared with stars or HI. Such simulations give insight on for example the pattern speed and pitch angle of the spiral arms, pattern speed and length of the bar, and how well a fixed spiral pattern versus live potential match the CO observations.  In some ways modelling external galaxies is easier since we have face on observations to compare with. For external galaxies, simulations have been able to successfully reproduce the large scale structure of M51 \citep{Salo2000,Salo2000b,Dobbs2010,Tress2020}, M33 \citep{Dobbs2018,Semczuk2018} and barred galaxies NGC4303 and NGC 3627 \citep{Iles2022}. Of those simulations which include molecular cloud analysis, \citet{Dobbs2019} are able to reproduce the radial variation of clouds in M33, and find that the molecular cloud properties depend primarily on gas surface density, and less on the spiral structure and stellar feedback.  Other studies similarly find that the molecular cloud properties are less dependent on interactions such as that of M51, and again primarily dependent on the gas density, and level of feedback \citep{Pettitt2020,Tress2020b}. An alternative approach by \citet{Sills2023} is to use observational gas densities and velocities as initial conditions, there for the Orion cloud, rather than a whole galaxy, although they had to make some assumptions about the cloud's properties to set up full 6D positions and velocities.

\section{Sub-galactic simulations or simulations resolving sub-cloud structure}
Simulations which span between the scales of clouds and galaxies have some key advantages compared to larger galaxy or smaller cloud scale simulations. It is difficult to well resolve clusters, or feedback processes over entire galaxy scales. Conversely, simulations on molecular cloud scales do not typically include larger galactic scale processes such as spiral arms, bars, larger supernovae bubbles, which may influence star formation.   
Galaxy scale simulations in general lack the resolution to resolve the detailed structure of molecular clouds, the full cloud population down to 1000 or 100 M$_{\odot}$, or clumps or cores within molecular clouds. A number of approaches have been made to incorporate galaxy scale effects simultaneously with high resolution. These include shearing box simulations, taking a section from a galaxy scale simulation and re-resolving at higher resolution, taking a vertical section of the galactic plane, and setting a higher level of resolution within a region of the galaxy disc. Shearing box simulations have the advantage that they can include spiral arms, and the gas is periodic, so such simulations can be run for a long period of time (10's or even 100's Myrs). It is difficult however to model for example a bar, or study the radial variation within a galaxy. The vertical section simulations have similar advantages and disadvantages, in that they can be run relatively long, but tend to only include variation in the z direction, so shear, spiral arms and  galactic rotation are not typically included. Simulations which remodel a section of a galaxy have the advantage that they can be relatively flexible, e.g. different regions can be selected, and it can be easy to include a potential or live stellar disc to allow the gas to follow the large scale dynamics. However such simulations can only be run for limited timescales due both to the boundary conditions and the increased computational cost at higher resolution. Allocating high resolution in a small part of the disc is the most satisfactory, as most of these disadvantages are not present, though the computational costs for including high resolution are still high so it is difficult to run the refined or high resolution simulations for long periods of time. For all these methods there are still concerns about the boundaries, or transition between different resolutions (e.g. \citealt{Seifried2017,Herrington2023}).

The TIGRESS simulations \citep{Kim2017,Kim2020}, building on previous work started by \citep{Kim2002}, model a shearing box of ISM with photoelectric heating,  supernovae and MHD. The inclusion of galactic rotation is found to be important for clouds to be sheared out and preventing large scale gravitational collapse. The simulations are run for around 700 Myr which allows some semi-equilibrium state to be reached within the region (after around 100 Myr).  The star formation rate is nevertheless cyclical, which leads to the emergence of galactic winds, which then fall back down on the disc \citep{Kim2018}. \citet{Kim2020} also include a spiral potential. The SILCC simulations model a vertical section of the disc, resolving the formation of molecular clouds, concentrating on molecular chemistry and the effects of feedback \citep{Walch2015}.

Other simulations have modelled a small region taken from a galaxy simulation \citep{VanLoo2013,Bonnell2013,Butler2015,Dobbs2015b,Smilgys2017,Bending2020,Dobbs2022a,Dobbs2022b,Bending2020,Herrington2023,Ali2023}. Some of the more recent of these papers focus more on clusters, following the evolution of clusters represented by groups of sink particles, but can also include initial conditions which have arisen from global galaxy simulations. \citet{Dobbs2022a} show that particularly massive clusters may emerge at the sites of strongly converging flows in galaxy, and that photoionisation is necessary to produced the observed radii of stellar clusters, whilst \citet{Dobbs2022b} show the emergence of OB associations in lower density regions of converging flows. Simulations on these scales also indicate that clusters, at least more massive clusters, appear to be formed hierarchically, through the mergers of smaller clusters \citep{Smilgys2017,Dobbs2022b}.

\citet{Smith2020} run a galaxy scale simulation and then modify the resolution within a 3 kpc box after the simulation reaches 150 Myr. This means they can continue modelling the whole galaxy, but simultaneously follow the formation and evolution of filamentary structures on small scales (Figure~\ref{fig:6}). They follow the evolution of filaments which are forming stars, whereby feedback breaks up the filaments, and quiescent filaments which undergo gravitational fragmentation. \citet{Renaud2013} also achieve very high resolution in their galaxy scale simulation, but in their case they re-resolve the whole galaxy simulation, again after the simulation has already run for about 240 Myr. 

\begin{figure}[h!]
\begin{center}
\includegraphics[width=18cm]{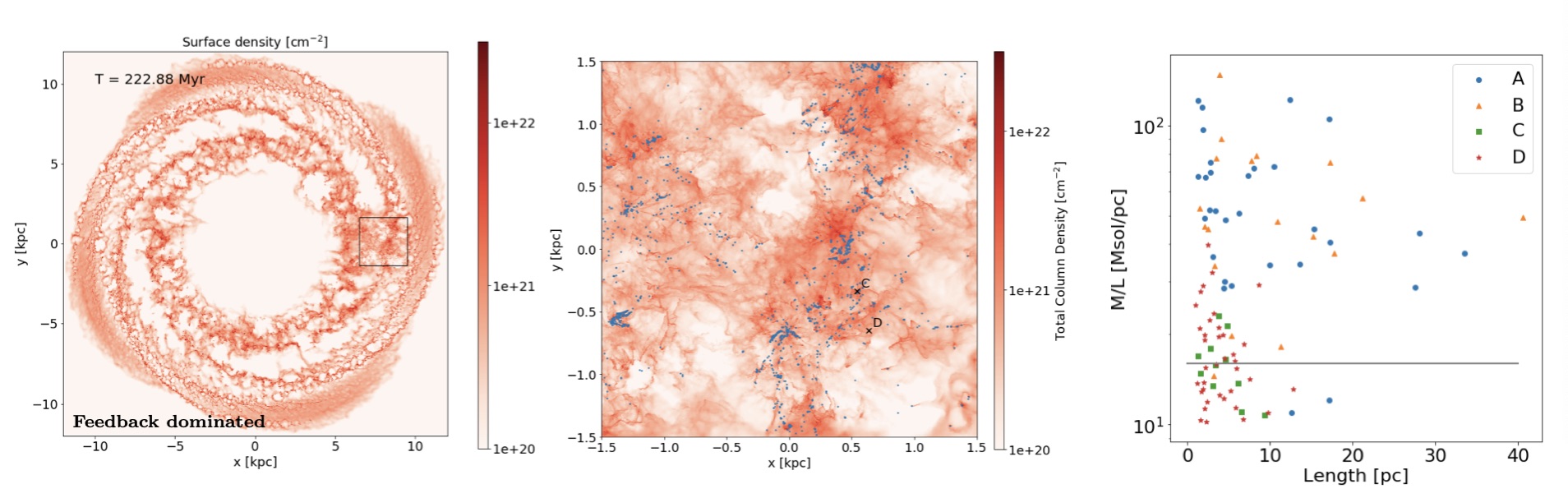}
\end{center}
\caption{A galaxy simulation is shown from \citet{Smith2020} (left), including supernovae feedback, and a zoom in high resolution section with higher refinement is shown in the middle panel, corresponding to the black box.  The blue dots in the middle panel represent sink particles, and the letters indicate selected cloud complexes. The right hand panel shows the distribution of filament mass to length ratios and lengths, the letters  C and D represent star forming regions dominated by feedback, whilst A and B represent regions dominated by the spiral arm potential where little feedback has occurred.}\label{fig:6}
\end{figure}

\section{Conclusions and looking ahead}
Simulations on isolated galaxy and subgalactic scales over the last 25 years or so have played a key role in developing our understanding of molecular cloud evolution and star formation within molecular clouds. 
Some of the earlier numerical work, carried out in the 2000s, focused on testing theories about instabilities in the ISM and molecular cloud formation that had been put forward in the previous decades, but which were unable to be investigated numerically until computers were sufficiently powerful to enable hydrodynamic simulations on kpc or galaxy scales. The simulations also took advantage of the fact that in the 1980s and 1990s the first observational surveys of molecular clouds gave statistical properties to compare with, and that the first surveys of clouds in external galaxies were starting to be made. 
With regards to molecular cloud formation, gravitational instabilities coupled with cooling, and cloud-cloud collisions are likely to be the most important processes in most galactic environments. Feedback is probably the main driver of the observed velocity dispersions in molecular clouds, though gravity and convergent flows or tidal velocity fields may also be significant, particularly in regions less impacted by massive stars.
 Cloud mass spectra, which are reproduced in most simulations unless they are dominated by gravitational collapse into massive clouds, can be explained theoretically either through gravitational instabilities in a multiphase medium \citep{Elmegreen1989}, or through cloud-cloud collisions \citep{Kwan1979}. Again probably both are present.

As the resolution of CO surveys has improved, observations have been able to probe both a larger range of external galaxies (e.g. PHANGS \citealt{Leroy2021}) and environments within galaxies. At the same time, simulations have investigated the role of spiral arms, bars and galactic centres. So far spiral arms are not found to have a large role in the properties of molecular clouds or galactic star formation rates, as found in both simulations and observations. Though spiral arms do appear to be able to produce more massive, and highly star forming GMCs. Bars may be more complex, as they host low density, low star forming regions, regions with strong shear channeling gas to the centre, as well as massive highly star forming regions. The last two of these have at least been investigated in simulations of bars and the Galactic Centre. So far there has been a considerable advance of our understanding of star formation at the Galactic Centre through simulations and observations. However it is not clear how typical our Galaxy Centre is compared to galaxies generally in the local universe. 

Since most surveys focus on the inner parts of galaxies, outer regions have not been such a focus of more recent work, however they may become increasingly significant if we are to understand the longer term evolution, and the role of accretion. Zoom in models of individual galaxies from cosmological simulations may be preferable in this case since they capture the surrounding environment. The LYRA simulations \citep{Gutcke2022} take initial conditions from the EAGLE simulations and achieve 4 M$_{\odot}$ per cell resolution but focus on dwarf galaxies. The VINTERGATAN zoom in simulations of the Milky Way \citep{Agertz2021} have a gas mass resolution of 7070 M$_{\odot}$, which is a slightly higher than isolated spiral galaxy simulations but unlike isolated galaxy models, these were run for the entirety of the Milky Way's history.

The past decade or so has seen much more interest in linking the smaller scale effects of feedback with the global evolution of clouds. Most simulations now find fairly short lifetimes, as based on the constituent gas present in the cloud, not the lifetime of H$_2$ molecules.  Most work on smaller scales indicates the role of photoionisation in immediately dispersing gas from clouds, in agreement with observations \citep{Chevance2020} although on larger galactic scales, it is not clear whether photoionisation, supernovae, or both equally drive the structure. The role of magnetic fields has been studied considerably in simulations of turbulence in the ISM, but more simulations are now starting to look at MHD on galaxy scales.

So far most simulations have investigated the properties of the gas, but a logical next step, resources and codes provided, would be to study the formation of clusters, OB associations and if applicable, isolated star formation. Following the full formation and evolution of clusters by resolving the formation of individual stars is still restricted to relatively small cloud scales, and achieving this on large, e.g. galaxy scales seems impractical for the foreseeable future. For example to resolve individual star formation down to brown dwarfs at the opacity limit with sink particles, \citet{Bate2003} resolve densities of 10$^{-13}$ g cm$^{-3}$, and use particle masses of $10^{-5}$ M$_{\odot}$. On a galactic scale modelling a fairly low mass of gas of $10^9$ M$_{\odot}$ with an SPH code would require $10^{14}$ SPH particles. Even if a single particle represented a solar or higher mass star this would still require 10 billion SPH particles. However there are still ways to more approximately model and follow cluster formation and evolution. One approach is to semi-evolve the clusters by adopting a subgrid model for their formation and dispersal. This approach has the advantage that it can be implemented on any scale simulation, on galaxy scales and cosmological scales \citep{Pfeffer2018,ReinaCampos2022,Grudic2023}. For example in \citet{ReinaCampos2022} each star particle formed in a star formation event then represents the fraction of star formation that ends up in a bound cluster. An alternative is simply to use sink or star particles \citep{Hennebelle2014,Gatto2017,Geen2018,Kim2018,Smith2020,Li2022,Dobbs2022a}, but each of these still represents a group of stars rather than a single star particle. 

An approach which potentially allows the full modelling of stellar clusters even with low resolution in the gas is to decouple the gas resolution from that of the stars. When star formation occurs, star particles are inserted with masses according to an assumed IMF \citep{Fujii2021,Rieder2022,Lahen2023}. For example \citet{Rieder2022} model the formation of clusters resolving all stars with masses down to 0.1 M$_{\odot}$ in a section of a spiral galaxy. \citet{Lahen2023} also model the full IMF of stars in simulations of dwarf galaxies. There are still drawbacks to this method, for example a decision still has to be made about the spatial and velocity distribution of the stars when they added. However reasonable estimates can be made given the spatial dimensions of the simulation \citep{Liow2022}.  Hybrid N-body$+$hydrodynamics codes \citep{Fujii2021b,Rieder2022} enable simulations to be more tractable with very large numbers of star particles, as quickly becomes the case once spatial scales become the size of clouds or larger.

\subsubsection{Permission to Reuse and Copyright}
Permission has not yet been sought to use the figures in the manuscript.

\section*{Conflict of Interest Statement}

The authors declare that the research was conducted in the absence of any commercial or financial relationships that could be construed as a potential conflict of interest.

\section*{Author Contributions}

The Author Contributions section is mandatory for all articles, including articles by sole authors. If an appropriate statement is not provided on submission, a standard one will be inserted during the production process. The Author Contributions statement must describe the contributions of individual authors referred to by their initials and, in doing so, all authors agree to be accountable for the content of the work. Please see  \href{https://www.frontiersin.org/guidelines/policies-and-publication-ethics#authorship-and-author-responsibilities}{here} for full authorship criteria.

\section*{Funding}
CLD is funded by the European Research Council H2020-EU.1.1 ICYBOB project (Grant No. 818940).

\section*{Acknowledgments}
This is a short text to acknowledge the contributions of specific colleagues, institutions, or agencies that aided the efforts of the authors.

\section*{Data Availability Statement}
The datasets [GENERATED/ANALYZED] for this study can be found in the [NAME OF REPOSITORY] [LINK].

\bibliographystyle{Frontiers-Harvard} 
\bibliography{Dobbs}




\end{document}